\documentclass[preprint,preprintnumbers,amsmath,amssymb,superscriptaddress]{revtex4}

\usepackage{graphicx}
\usepackage{dcolumn}
\usepackage{bm}

\makeatletter
\def\@seccntformat#1{}
\makeatother
\renewcommand{\numberline}[1]{}

\begin{document}

Theoretical and experimental discovery of single-Dirac-cone topological-insulator class was reported in the same paper at arXiv:0812.2078 (2008) [http://arxiv.org/abs/0812.2078]. Perspectives and Published version at Y. Xia et.al., Nature Physics 5, 398-402 (2009) [http://dx.doi.org/10.1038/nphys1294].



\title{First observation of Spin-Momentum helical locking in Bi$_2$Se$_3$ and Bi$_2$Te$_3$, demonstration of Topological-Order at 300K and a realization of topological-transport-regime}

\author{D. Hsieh}
\affiliation{Joseph Henry Laboratories of Physics, Department of Physics, Princeton
University, Princeton, NJ 08544, USA}

\author{Y. Xia}
\affiliation{Joseph Henry Laboratories of Physics, Department of Physics, Princeton
University, Princeton, NJ 08544, USA}

\author{D. Qian}
\affiliation{Joseph Henry Laboratories of Physics, Department of
Physics, Princeton University, Princeton, NJ 08544, USA}
\affiliation{Department of Physics, Shanghai Jiao Tong University,
Shanghai 200030, China}

\author{L.A. Wray}
\affiliation{Joseph Henry Laboratories of Physics, Department of
Physics, Princeton University, Princeton, NJ 08544, USA}

\author{J. H. Dil}
\affiliation{Swiss Light Source, Paul Scherrer Institute, CH-5232,
Villigen, Switzerland} \affiliation{Physik-Institut, Universit\"{a}t
Z\"{u}rich-Irchel, 8057 Z\"{u}rich, Switzerland}

\author{F. Meier}
\affiliation{Swiss Light Source, Paul Scherrer Institute, CH-5232,
Villigen, Switzerland} \affiliation{Physik-Institut, Universit\"{a}t
Z\"{u}rich-Irchel, 8057 Z\"{u}rich, Switzerland}

\author{J. Osterwalder}
\affiliation{Physik-Institut, Universit\"{a}t Z\"{u}rich-Irchel,
8057 Z\"{u}rich, Switzerland}

\author{L. Patthey}
\affiliation{Swiss Light Source, Paul Scherrer Institute, CH-5232,
Villigen, Switzerland}

\author{J. G. Checkelsky}
\affiliation{Joseph Henry Laboratories of Physics, Department of
Physics, Princeton University, Princeton, NJ 08544, USA}

\author{N. P. Ong}
\affiliation{Joseph Henry Laboratories of Physics, Department of
Physics, Princeton University, Princeton, NJ 08544, USA}

\author{A. V. Fedorov}
\affiliation{Advanced Light Source, Lawrence Berkeley Laboratory,
Berkeley, CA 94720, USA}

\author{H. Lin}
\affiliation{Department of Physics, Northeastern University, Boston,
MA 02115, USA}

\author{A. Bansil}
\affiliation{Department of Physics, Northeastern University, Boston,
MA 02115, USA}

\author{D. Grauer}
\affiliation{Department of Chemistry, Princeton University,
Princeton, NJ 08544, USA}

\author{Y. S. Hor}
\affiliation{Department of Chemistry, Princeton University,
Princeton, NJ 08544, USA}

\author{R. J. Cava}
\affiliation{Department of Chemistry, Princeton University,
Princeton, NJ 08544, USA}

\author{M. Z. Hasan}
\affiliation{Joseph Henry Laboratories of Physics, Department of
Physics, Princeton University, Princeton, NJ 08544, USA}
\affiliation{Princeton Center for Complex Materials, Princeton
University, Princeton NJ 08544, USA} \email{mzhasan@Princeton.edu}




\maketitle

\textbf{Helical Dirac fermions - charge carriers that behave as
massless relativistic particles with a quantum magnetic moment that
is locked to its direction of motion - are proposed to be the key to
realizing fundamentally new phenomena in condensed matter physics
\cite{1,2}. Prominent examples include the anomalous quantization of
magneto-electric coupling \cite{3,4,5,6}, half-fermion states that
are their own anti-particle \cite{7,8}, and charge fractionalization
in a Bose-Einstein condensate \cite{9}, all of which are not
possible with conventional Dirac fermions of the graphene variety
\cite{10}. Helical Dirac fermions are challenging to find, and so
far remain undiscovered, because they are forbidden to exist in
ordinary Dirac materials such as graphene \cite{10} or bismuth
\cite{11}, and because the necessary spin-sensitive measurements are
lacking. It has recently been proposed that helical Dirac fermions
may exist at the edges of certain types of topologically ordered
insulators \cite{3,4,12}, and that their peculiar properties may be
accessed provided the insulator is tuned into a topological
transport regime \cite{3,4,5,6,7,8,9}. However, helical Dirac
fermions have not been observed in existing topological insulators
\cite{13,14,15,16,17,18} to date because conventional electrical
gating based tuning techniques that work for graphene \cite{10,19}
cannot be applied and the necessary spin-resolved photoelectron
detections do not exist. Here we report the first realization of a
tunable topological insulator in a bismuth based class by combining
spin- and momentum-resolved spectroscopies, bulk charge compensation
and surface control. Our results reveal a spin-momentum locked Dirac
cone that is nearly 100\% spin-polarized, which exhibits a tunable
topological fermion density in the vicinity of the Kramers' point
and can be driven to the long-sought topological transport regime.
The observed topological nodal Dirac ground state is found to be
protected even up to room temperature (300 K). Our results on
Bi$_{2-\delta}$Ca$_{\delta}$Se$_3$.NO$_2$ pave the way for future
transport based studies of topological insulators, and possible room
temperature applications of protected spin-polarized edge channels
in microelectronics technology.}


Unlike conventional Dirac fermions as in graphene, helical Dirac
fermions possess a net spin and are guaranteed to be conducting
because of time-reversal symmetry \cite{3,4,5}, allowing the unique
possibility to carry spin currents without heat dissipation.
However, the most important difference lies in the topological
properties of helical Dirac fermion systems \cite{3,4,5,12}, which
are predicted to manifest in several ways, provided the system can
be tuned to the topological transport regime where the charge
density vanishes (analogous to the charge neutrality point in
graphene \cite{19}). These manifestations include an anomalous
half-integer quantization of Hall conductance \cite{3,4,5,6}, a
realization of Majorana fermions (particles with anyon exchange
statistics that differs from the conventional Bose or Fermi-Dirac
statistics) \cite{7,8}, and a generation of fractionally charged
particles \cite{9}. Helical fermions are believed to exist on the
edges of certain types of three-dimensional (3D) topological
insulators \cite{3,4,12}, with material candidates Bi$_2$X$_3$ (X =
Se, Te) recently proposed based on observations \cite{15,17} and
models \cite{4,20}. However, these materials cannot be used to
detect helical Dirac fermion physics for three reasons: First, the
helical properties of the surface electrons are unknown and depend
on the materials class. Second, their electronic structure is not in
the topological transport regime, thus not allowing any of the
interesting topological insulator experiments to be performed to
this date. And third, unlike 2D quantum Hall Dirac systems such as
graphene \cite{10,19}, 3D topological insulators cannot be tuned to
this zero carrier density regime through electrical gating, which
has prevented a revolution like that witnessed for graphene
\cite{10} from taking place for topological insulators.

\begin{figure}
\includegraphics[scale=0.6,clip=true, viewport=-0.0in 0in 11.0in 5.5in]{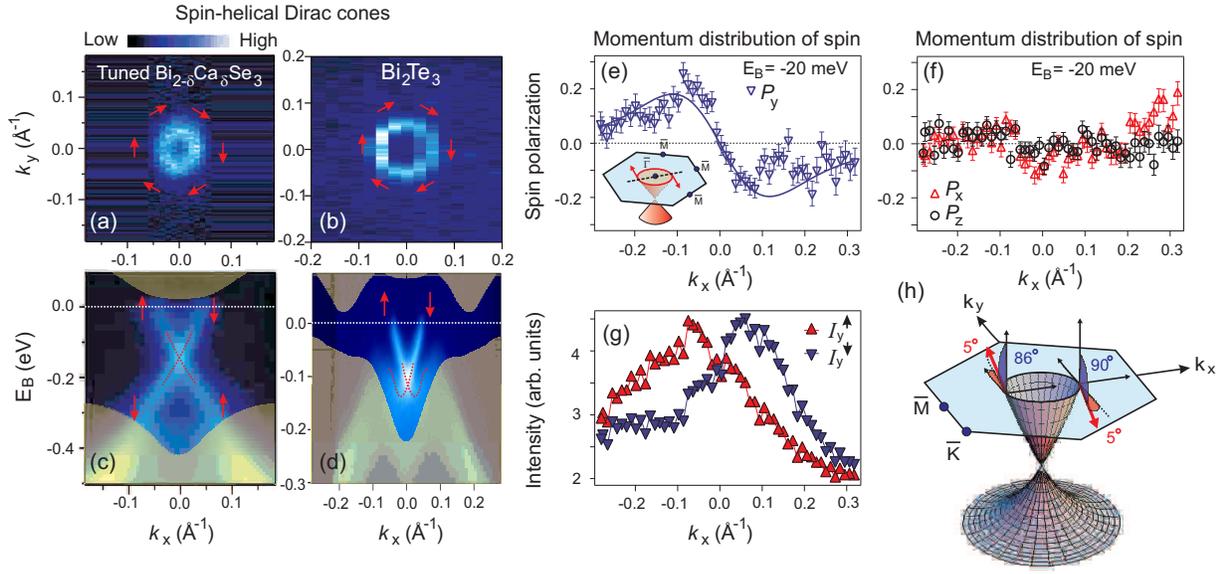}
\caption{\label{fig:Fig1} Hsieh, Xia, Qian $et$ $al.$, NATURE 460,
1101 (2009).}
\end{figure}

To determine the key helical properties of the surface electrons
near the Fermi level ($E_F$) in the candidate Bi$_2$X$_3$ class, we
performed spin- and angle-resolved photoemission spectroscopy
(ARPES) scans using a double Mott detector setup \cite{21}, which
systematically measures all three components of the spin of a
photoelectron as a function of its energy and momentum throughout
the Brillouin zone (Supplementary Information). Although the surface
electrons of both Bi$_2$Se$_3$ and Bi$_2$Te$_3$ exhibit a finite
density of states near $E_F$ (Figs 1(a)-(d)), there is an additional
contribution to the density of states around momentum $\bar{\Gamma}$
from the spin-degenerate bulk conduction band in Bi$_2$Se$_3$
\cite{15}. Therefore, the helical nature of the surface electrons is
most clearly resolved in Bi$_2$Te$_3$. We analyzed the
spin-polarization of photoelectrons emitted at a binding energy
$E_B$ = -20 meV along the $k_x$ ($\parallel$ $\bar{\Gamma}$-\={M})
cut in Bi$_2$Te$_3$ (inset Fig. 1(e)). Because the surface state
dispersion of Bi$_2$X$_3$ exhibits a pronounced time dependence
after cleavage (SI) related to semiconductor band bending effects
\cite{17}, data collection times were only long enough to ensure a
level of statistics sufficient to measure the spin-polarized
character of the surface states.

Figures 1(e) and (f) show the measured spin polarization spectra
$P_i$ of the $i$ = $x$, $y$ and $z$ (out-of-plane) components along
the $\bar{\Gamma}$-\={M} direction. In the $x$ and $z$ directions,
no clear signal can be discerned within the margins of statistical
error. In the $y$ direction on the other hand, clear polarization
signals of equal magnitude and opposite sign are observed for
surface electrons of opposite momentum, evidence that the spin and
momentum directions are one-to-one locked. This is most clearly seen
in the spin-resolved spectra ($I_y^{\uparrow,\downarrow}$; Fig.
1(g)), which are calculated from $P_y$ according to $I_y$  =
$I_{tot}$(1+$P_y$)/2 and $I_y$ = $I_{tot}$(1-$P_y$)/2, where
$I_{tot}$ is the spin-averaged intensity. To extract the spin
polarization vectors of the forward (+$k_x$) and backward (-$k_x$)
moving electrons, we performed a standard numerical fit (SI)
\cite{21}. The fit results yield 100($\pm$15)\% polarized (see Fig.
1(h)) spins that point along the (\textbf{k}$\times$\textbf{z})
direction, which is consistent with its spin-orbit coupling origin
\cite{14,21}. Our combined observations of a linear dispersion
relation and a one-to-one locking of momentum and spin directions
allow us to conclude that the surface electrons of Bi$_2$X$_3$ (X =
Se, Te) are helical Dirac fermions of $Z_2$ topological order origin
(Fig. 1).

\begin{figure}
\includegraphics[scale=0.65,clip=true, viewport=0.1in 0in 11.3in 6.0in]{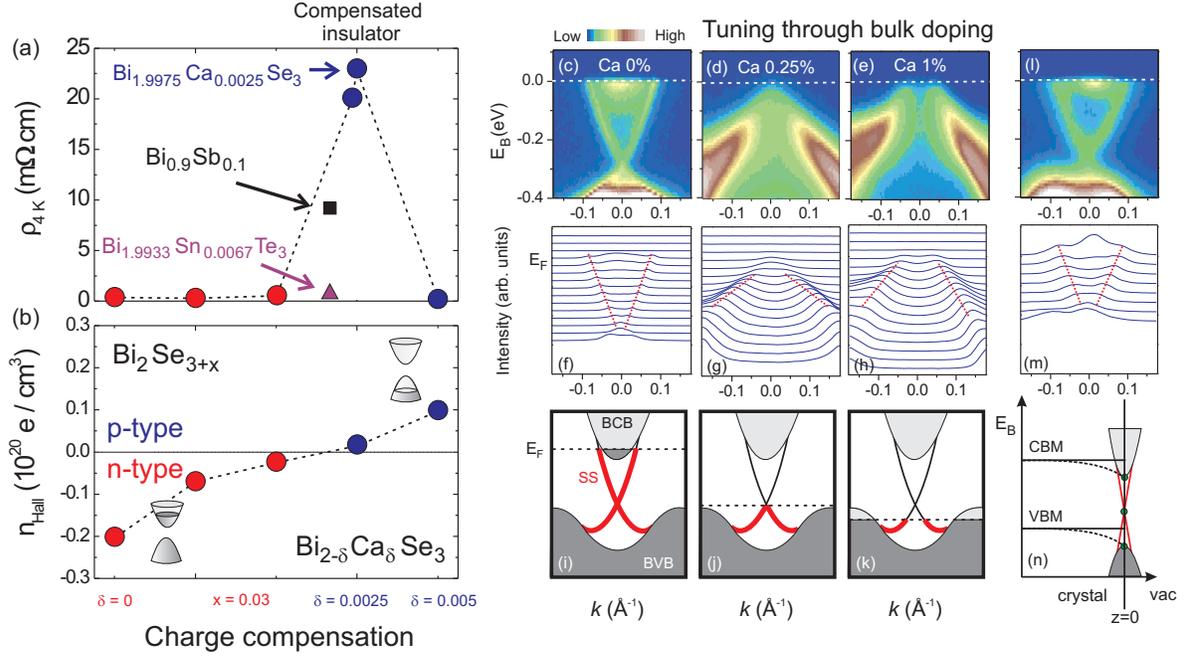}
\caption{\label{fig:Fig2} Hsieh, Xia, Qian $et$ $al.$, NATURE 460,
1101 (2009).}
\end{figure}

To experimentally access these helical Dirac fermions for research
device applications, the electronic structure must be in the
topological transport regime where there is zero charge fermion
density \cite{7,8,9}. This regime occurs when $E_F$ lies in between
the bulk valence band maximum (VBM) and the bulk conduction band
minimum (CBM), and exactly at the surface Dirac point, which should
in turn lie at a Kramers' time-reversal invariant momentum
\cite{3,4}. This is clearly not the case in either Bi$_2$Te$_3$,
Bi$_2$(Sn)Te$_3$, Bi$_2$Se$_3$ or graphene. Although pure
Bi$_2$X$_3$ are expected to be undoped semiconductors
\cite{20,22,23}, nominally stoichiometric samples are well known to
be n- and p-type semiconductors due to excess carriers introduced
via Se or Te site defects respectively \cite{16,17}. To compensate
for the unwanted defect dopants, trace amounts of carriers of the
opposite sign must be added into the naturally occurring material,
which may be easier to achieve in Bi$_2$Se$_3$ than in Bi$_2$Te$_3$
because it has a larger band gap \cite{24} (around 0.35 eV \cite{25}
compared to 0.18 eV \cite{26} respectively). To lower the $E_F$ of
Bi$_2$Se$_3$ into the bulk band gap, we substituted trace amounts of
Ca$^{2+}$ for Bi$^{3+}$ in as-grown Bi$_2$Se$_3$, where Ca has been
shown to act as a hole donor by scanning tunnelling microscopy and
thermoelectric transport studies \cite{16}. Figure 2(a) shows that
as the Ca concentration increases from 0\% to 0.5\%, the low
temperature resistivity sharply peaks at 0.25\%, which suggests that
the system undergoes a metal to insulator to metal transition. The
resistivity peak occurs at a Ca concentration where a change in sign
of the Hall carrier density also is observed (Fig. 2(b)), which
shows that for measured Ca concentrations below and above 0.25\%,
electrical conduction is supported by electron and hole carriers
respectively.

\begin{figure}
\includegraphics[scale=0.65,clip=true, viewport=-0.0in 0in 6.8in 6.8in]{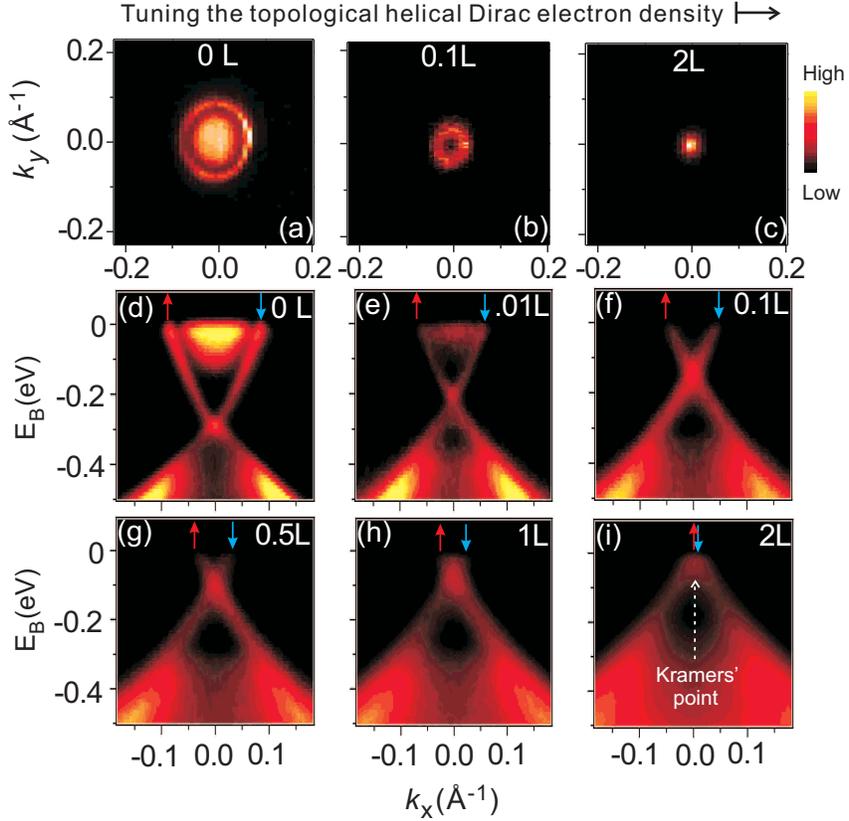}
\caption{\label{fig:Fig3} Hsieh, Xia, Qian $et$ $al.$, NATURE 460,
1101 (2009).}
\end{figure}

We performed systematic time-dependent ARPES measurements to study
the electronic structure evolution of
Bi$_{2-\delta}$Ca$_{\delta}$Se$_3$ as a function of Ca doping in
order to gain insight into the trends observed in transport (Figs
2(a) and (b)). Early time ARPES energy dispersion maps taken through
the $\bar{\Gamma}$ point of the (111) surface Brillouin zone are
displayed in Figure 2(c)-(h) for several Ca doping levels. In the
as-grown ($\delta$ = 0) Bi$_2$Se$_3$ samples, a single surface Dirac
cone is observed with $E_F$ lying nearly 0.3 eV above the Dirac node
forming an electron Fermi surface (FS). We also observe that $E_F$
intersects the electron-like bulk conduction band. When a 0.25\%
concentration of Ca is introduced, $E_F$ is dramatically lowered to
lie near the Dirac node (Fig. 2(d)), which is consistent with Ca
acting as a highly effective hole donor. Because the bulk CBM lies
at a binding energy of approximately -0.1 eV for $\delta$ = 0 (Fig.
2(c)), an 0.3 eV shift in $E_F$ between $\delta$ = 0 and $\delta$ =
0.0025 suggests that for $\delta$ = 0.0025, $E_F$ is located 0.2 eV
below the CBM. This is consistent with $E_F$ being in the bulk band
gap because the indirect energy gap between the CBM and the VBM is
known from both tunneling \cite{24} and optical \cite{25} data and
theory \cite{22} to be nearly 0.35 eV.

As the Ca concentration is increased further, the position of $E_F$
continues a downward trend such that by $\delta$ = 0.01, it is
located clearly below the Dirac node (Fig. 2(e)) and intersects the
hole-like bulk valence band. The systematic lowering of $E_F$ with
increasing $\delta$ in Bi$_{2-\delta}$Ca$_{\delta}$Se$_3$ observed
in early time ARPES measurements (Figs 2(i)-(k)), which reflect the
electronic structure of the sample bulk, consistently explain the
measured transport behavior. However, we observe that $E_F$ rises
back up over time across all samples such that all spectra relax
back to a $\delta$ = 0 like spectrum on a typical time scale of 18
hours (Fig. 2(l)). Such a slow upward shift of the surface Fermi
level has been observed also in Bi$_2$Te$_3$ \cite{17} and is due to
a surface band bending effect commonly observed in most
semiconductors (SI). Therefore, although bulk Ca doping succeeds in
tuning $E_F$ between the bulk valence and conduction bands, it does
not change the position of $E_F$ relative to the surface Dirac point
in the ground state.

Because the surface Dirac point in the ground state of most
insulating Bi$_{1.9975}$Ca$_{0.0025}$Se$_3$ lies $\sim$0.3 eV below
$E_F$, its electronic structure is still not in the much desired
topological transport regime. To bring the surface Dirac point level
with $E_F$ in Bi$_{2-\delta}$Ca$_{\delta}$Se$_3$, we show that hole
carriers can be systematically introduced into the surface by dosing
with NO$_2$ molecules, which has been demonstrated in graphene
\cite{27,28}. Figure 3 shows that with increasing surface hole donor
concentration, the binding energy of the surface Dirac point rises
monotonically towards $E_F$. Starting from $E_B$ $\sim$ -0.3 eV at a
0 Langmuir (L) dose, it rises to -0.15 eV at 0.1 L where the surface
bent CBM has completely disappeared, and finally to the charge
neutrality point ($E_B$ = 0 eV) at 2 L. No further changes of the
chemical potential are observed with higher dosages. To quantify the
surface carrier density ($n$) dependence on surface hole donor
concentration, we mapped the surface state FS in Figures 3(a)-(c)
and performed an electron count based on FS area $n$ =
$A_{FS}$/$A_{BZ}$, where $A_{FS}$ is the area of the FS and $A_{BZ}$
is the area of the surface Brillouin zone. We find that 0.1 L of
NO$_2$ removes approximately 0.0066 electrons per surface unit cell
of Bi$_{2-\delta}$Ca$_{\delta}$Se$_3$ (111), and an excess of 2 L
reduces the FS to a single point within our experimental resolution,
which has an additional 0.005 electrons per unit cell removed.
Because surface doping does not affect the carrier density in the
bulk (which thus remains insulating), the energy of the Dirac point
is lifted above the bulk VBM; a new time independent electronic
ground state is realized that lies in the topological transport
regime with $E_F$ intersecting the Dirac node.

In order to investigate the thermal stability of this nodal Dirac
ground state (Fig. 4(e)), temperature dependent ARPES scans were
collected on Bi$_{2-\delta}$Ca$_{\delta}$Se$_3$ samples that were
first surface hole doped with NO$_2$ at a temperature $T$ = 10 K.
Figures 4(c) and (d) illustrate that the charge neutral point-like
FS (Fig. 4(a)) is robust up to room temperature ($T$ = 300 K) over
days long measurement times. A density of states that decreases
linearly to zero at the Dirac point energy at 300 K (Fig. 4(f)) is
further evidence that the low energy properties of
Bi$_{1.9975}$Ca$_{0.0025}$Se$_3$.NO$_2$ are dominated by a novel
ground state that features massless helical Dirac fermions with
nearly 100\% spin polarization. This also confirms a non-trivial
$\pi$ Berry's phase on the surface due to the spin-momentum locking
pattern that we observe, which is similar to the robust Berry's
phase previously observed in the Bi-Sb system \cite{14} (Fig. 1).

Helical nodal Dirac fermions are forbidden from acquiring a mass
through band gap formation because they are located around
time-reversal invariant (Kramers') momenta $k_T$ = $\bar{\Gamma}$ or
\={M} (Fig. 4(h)). This makes them fundamentally different from
chiral Dirac fermions such as those found in graphene, which are
located at \={K} and not topologically protected (Fig. 4(g)). The
helical Dirac fermion on the surface of Bi$_2$Se$_3$ owes its
existence to a non-zero topological number $\nu_0$ given by
(-1)$^{\nu_0}$ = $\Pi_{k_T}\Pi_{m=1}^{N}\xi_{2m}$($k_T$), where
$\xi_{2m}$($k_T$) is the parity eigenvalue of the bulk wavefunction
at the 3D Kramers' point $k_T$ and $N$ is the number of occupied
bulk bands \cite{4}. Because Ca dopants are present in only trace
quantities in Bi$_{1.9975}$Ca$_{0.0025}$Se$_3$.NO$_2$, the values of
$\xi_{2m}$($k_T$) do not deviate from those of pure Bi$_2$Se$_3$, as
evidenced by the persistence of a single gapless surface band in
both Bi$_2$Se$_3$ and Bi$_{1.9975}$Ca$_{0.0025}$Se$_3$. Both
Ca$^{2+}$ and NO$^{2-}$ are non-magnetic and so do not break
time-reversal symmetry, therefore the same topological number
($\nu_0$ = 1) applies even in the Dirac transport regime (Fig. 4)
realized by our method shown here, which is stable with both time
and temperature. Our direct demonstration of spin-polarized edge
channels and room temperature operability of chemically gated
stoichiometric Bi$_2$Se$_3$ or Bi$_{2-\delta}$Ca$_{\delta}$Se$_3$,
not achieved in purely 2D topological systems such as Hg(Cd)Te
quantum wells \cite{29}, enables exciting future room temperature
experiments on surface helical Dirac fermions that carry a
non-trivial $\pi$ Berry's phase.

\begin{figure}
\includegraphics[scale=0.55,clip=true, viewport=-0.0in 0in 11.2in 6.0in]{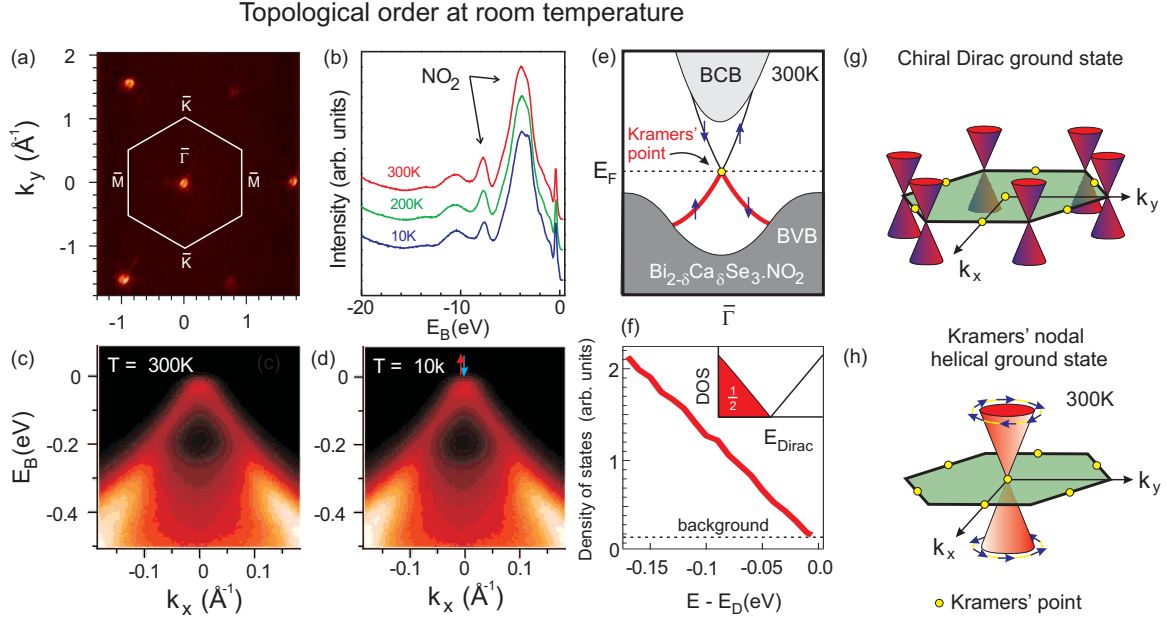}
\caption{\label{fig:Fig4} Hsieh, Xia, Qian $et$ $al.$, NATURE 460,
1101 (2009).}
\end{figure}

Our demonstration of topological order at room temperature opens up
possibilities of using quantum Hall-like phenomena and
spin-polarized protected edge channels for spintronic or computing
device applications without the traditional requirements of high
magnetic fields and delicate cryogenics. A direct detection of
surface-edge states would be possible in stoichiometric Bi$_2$Se$_3$
or Bi$_{2-\delta}$Ca$_{\delta}$Se$_3$, using transport methods which
will bear signatures of weak anti-localization and thus exhibit
anomalous magneto-optic effects. Here we envisage a few sample
experiments that could be carried out by using surface doped or
electrically gated Bi$_{2-\delta}$Ca$_{\delta}$Se$_3$. By applying a
weak time-reversal breaking perturbation at the surface of
Bi$_{2-\delta}$Ca$_{\delta}$Se$_3$.NO$_2$ so as to lift the Kramers'
degeneracy at $E_F$ (A method of gap opening on the surface is shown
in the SI), a half-integer quantized magneto-electric coupling can
be realized \cite{3,4,5,6}, which may be measured by standard Hall
probes. This would enable a variety of novel surface quantum Hall
physics to be realized. Another class of experiments would be made
possible by interfacing the helical topological surface with
magnetic and ordinary superconducting films. An interferometer
device can be built based on Bi$_{2-\delta}$Ca$_{\delta}$Se$_3$ that
is used to create and detect long-sought Majorana fermions
\cite{7,8}. These particles, which have never been experimentally
observed, possess only half the degrees of freedom of a conventional
fermion and constitute the key building block for topological
quantum computing that can operate in fault-tolerant mode. Yet
another class of experiments would be made possible by sandwiching a
charge neutral topological insulator film made of
Bi$_{2-\delta}$Ca$_{\delta}$Se$_3$ within a charged capacitor. In
this way, a microchip that supports a topological electron-hole
condensate with fractionally charged vortices \cite{9} can be
fabricated, which offers the exciting opportunity to probe
interactions between Dirac fermions of opposite helicity; this would
enable searching for exotic quantum phenomena beyond the standard
model of particle physics \cite{30}.

\vspace{1cm}

\small{\textbf{Acknowledgements} The spin-resolved ARPES
measurements are supported by NSF through the Center for Complex
Materials (DMR-0819860) and Princeton University; the use of
synchrotron X-ray facilities (ALS-LBNL Berkeley) is supported by the
Basic Energy Sciences of the U.S. Department of Energy
(DE-FG-02-05ER46200) and by the Swiss Light Source, Paul Scherrer
Institute, Villigen, Switzerland.}

\vspace{1cm}

\small{\textbf{Author information} Correspondence and requests for
materials should be addressed to M.Z.H (mzhasan@princeton.edu).}

\section{Methods}

Spin-integrated ARPES data were taken at beamlines 12.0.1 and 7.0.1
of the Advanced Light Source in Lawrence Berkeley National
Laboratory with 29 eV to 100 eV photons. Typical energy and momentum
resolutions were 15 meV and 1\% of the surface BZ and 50 meV and 2\%
of the surface BZ respectively. Spin-resolved ARPES measurements
were performed at the SIS beamline at the Swiss Light Source using
the COPHEE spectrometer, which consists of two 40 kV classical Mott
detectors that measure all three spatial components of spin
polarization. Spin-resolved measurements were taken with 20 eV to 22
eV photons with energy and momentum resolutions of 80 meV and 3\% of
the surface BZ respectively. Spin-integrated data were collected on
Bi$_{2-\delta}$Ca$_{\delta}$Se$_3$ and Bi$_2$Te$_3$ single crystals
cleaved in ultra high vacuum pressures better than $5\times10^{-11}$
torr and maintained at a temperature of 10 K unless otherwise
specified. Spin-resolved data were collected at 50 K. Adsorption of
NO$_2$ molecules on Bi$_{2-\delta}$Ca$_{\delta}$Se$_3$ was achieved
via controlled exposures to NO$_2$ gas (Matheson, 99.5\%). The
adsorption effects were studied under static flow mode by exposing
the cleaved sample surface to the gas for a certain time then taking
data after the chamber was pumped down to the base pressure. Spectra
of the NO$_2$ adsorbed surfaces were taken within minutes of opening
the photon shutter to minimize photon exposure related effects. The
calculations were performed with the LAPW method in slab geometry
using the WIEN2K package.


\begin{thebibliography}{h}

\bibitem{1}
Day, C. Exotic spin textures show up in diverse materials.
\textit{Phys. Today} \textbf{62}, 12-13 (2009).
http://dx.doi.org/10.1063/1.3120883 

\bibitem{2}
Moore, J. E. Topological insulators: The next generation.
\textit{Nature Phys.} \textbf{5}, 378-380 (2009).
http://dx.doi.org/10.1038/nphys1294

\bibitem{3}
Fu, L., Kane, C. L. \& Mele, E. J. Topological insulators in three
dimensions. \textit{Phys. Rev. Lett.} \textbf{98}, 106803 (2007).

\bibitem{4}
Fu, L. \& Kane, C. L. Topological insulators with inversion
symmetry. \textit{Phys. Rev. B.} \textbf{76}, 045302 (2007).

\bibitem{5}
Qi, X.-L., Hughes, T. L. \& Zhang, S.-C. Topological field theory of
time-reversal invariant insulators. \textit{Phys. Rev. B}
\textbf{78}, 195424 (2008).

\bibitem{6}
Essin, A., Moore, J. E. \& Vanderbilt, D. Magnetoelectric
polarizability and axion electrodynamics in crystalline insulators.
\textit{Phys. Rev. Lett.} \textbf{102}, 146805 (2009).

\bibitem{7}
Fu, L. \& Kane, C. L. Probing neutral Majorana fermion edge modes
with charge transport. \textit{Phys. Rev. Lett.} \textbf{102},
216403 (2009).

\bibitem{8}
Akhmerov, A. R., Nilsson, J. \& Beenakker, C. W. J. Electrically
detected interferometry of Majorana fermions in a topological
insulator. \textit{Phys. Rev. Lett.} \textbf{102}, 216404 (2009).

\bibitem{9}
Seradjeh, B., Moore, J. E. \& Franz, M. Exciton condensation and
charge fractionalization in a topological insulator film.
\textit{Phys. Rev. Lett.} \textbf{103}, 066402 (2009).

\bibitem{10}
Geim, A. K. \& Novoselov, K. S. The rise of graphene. \textit{Nature
Mat.} \textbf{6}, 183-191 (2007).

\bibitem{11}
Li, L. et al. Phase transitions of Dirac electrons in Bismuth.
\textit{Science} \textbf{321}, 547-550 (2008).

\bibitem{12}
Moore, J. E. \& Balents, L. Topological invariants of
time-reversal-invariant band structures. \textit{Phys. Rev. B}
\textbf{75} 121306(R) (2007).

\bibitem{13}
Hsieh, D. et al. A topological Dirac insulator in a quantum spin
Hall phase. \textit{Nature} \textbf{452}, 970-974 (2008).

\bibitem{14}
Hsieh, D. et al. Observation of unconventional quantum spin textures
in topological insulators. \textit{Science} \textbf{323}, 919-922
(2009).

\bibitem{15}
Xia, Y. et al. Observation of a large-gap topological-insulator
class with a single Dirac cone on the surface. \textit{Nature Phys.}
\textbf{5}, 398-402 (2009).

\bibitem{16}
Hor, Y. S. et al. p-type Bi$_2$Se$_3$ for topological insulator and
low-temperature thermoelectric applications. \textit{Phys. Rev. B}
\textbf{79}, 195208 (2009).

\bibitem{17}
Noh, H.-J. et al. Spin-orbit interaction effect in the electronic
structure of Bi$_2$Te$_3$ observed by angle-resolved photoemission
spectroscopy. \textit{Europhys. Lett.} \textbf{81}, 57006 (2008).

\bibitem{18}
Nishide, A. et al. Direct mapping of the spin-filtered surface bands
of a three-dimensional quantum spin Hall insulator. Preprint at
<http://arxiv.org/abs/0902.2251> (2009).

\bibitem{19}
Checkelsky, J. G., Li, L. \& Ong, N. P. Divergent resistance of the
Dirac point in graphene: Evidence for a transition in high magnetic
field. \textit{Phys. Rev. B} \textbf{79}, 115434 (2009).

\bibitem{20}
Zhang, H. et al. Topological insulators in Bi$_2$Se$_3$,
Bi$_2$Te$_3$ and Sb$_2$Te$_3$ with a single Dirac cone on the
surface. \textit{Nature Phys.} \textbf{5}, 438-442 (2009).

\bibitem{21}
Meier, F., Dil, J. H., Lobo-Checa, J., Patthey, L. \& Osterwalder,
J. Quantitative vectorial spin analysis in angle-resolved
photoemission: Bi/Ag(111) and Pb/Ag(111). \textit{Phys. Rev. B}
\textbf{77}, 165431 (2008).

\bibitem{22}
Larson, P. et al. Electronic structure of Bi$_2$X$_3$ (X = S, Se, T)
compounds: Comparison of theoretical calculations with photoemission
studies. \textit{Phys. Rev. B} \textbf{65}, 085108 (2001).

\bibitem{23}
Mishra, S. K., Satpathy, S. \& Jepsen, O. Electronic structure and
thermoelectric properties of bismuth telluride and bismuth selenide.
\textit{J. Phys: Condens. Mat.} \textbf{9}, 461-470 (1997).

\bibitem{24}
Urazhdin, S. et al. Surface effects in layered semiconductors
Bi$_2$Se$_3$ and Bi$_2$Te$_3$. \textit{Phys. Rev. B} \textbf{69},
085313 (2004).

\bibitem{25}
Black, J., Conwell, E. M., Seigle, L. \& Spencer, C. W. Electrical
and optical properties of some M$_2^{V-B}$N$_3^{VI-B}$
semiconductors. \textit{J. Phys. Chem. Sol.} \textbf{2}, 240-251
(1957).

\bibitem{26}
Thomas, G. A. et al. Large electron-density increase on cooling a
layered metal: Doped Bi$_2$Te$_3$. \textit{Phys. Rev. B}
\textbf{46}, 1553-1556 (1992).

\bibitem{27}
Zhou, S. et al. A. Metal to insulator transition in epitaxial
graphene induced by molecular doping. \textit{Phys. Rev. Lett.}
\textbf{101}, 086402 (2008).

\bibitem{28}
Schedin, F. et al. Detection of individual gas molecules adsorbed on
graphene. \textit{Nature Mat.} \textbf{6}, 652-655 (2007).

\bibitem{29}
Konig, M. et al. Quantum spin Hall insulator state in HgTe quantum
wells. \textit{Science} \textbf{318}, 766-770 (2007).

\bibitem{30}
Wilczek, F. Remarks on dyons. \textit{Phys. Rev. Lett.} \textbf{48},
1146-1149 (1982).

\end{thebibliography}

\newpage

\textbf{Figure 1 : Detection of spin-momentum locking of
spin-helical Dirac electrons in Bi$_2$Se$_3$ and Bi$_2$Te$_3$ using
spin-resolved ARPES.} (a) ARPES intensity map at $E_F$ of the (111)
surface of tuned Bi$_{2-\delta}$Ca$_{\delta}$Se$_3$ (see text) and
(b) the (111) surface of Bi$_2$Te$_3$. Red arrows denote the
direction of spin around the Fermi surface. (c) ARPES dispersion of
tuned Bi$_{2-\delta}$Ca$_{\delta}$Se$_3$ and (d) Bi$_2$Te$_3$ along
the $k_x$ cut. The dotted red lines are guides to the eye. The
shaded regions in (c) and (d) are our calculated projections of the
bulk bands of pure Bi$_2$Se$_3$ and Bi$_2$Te$_3$ onto the (111)
surface respectively. (e) Measured $y$ component of
spin-polarization along the $\bar{\Gamma}$-\={M} direction at $E_B$
= -20 meV, which only cuts through the surface states. Inset shows a
schematic of the cut direction. (f) Measured $x$ (red triangles) and
$z$ (black circles) components of spin-polarization along the
$\bar{\Gamma}$-\={M} direction at $E_B$ = -20 meV. Error bars in (e)
and (f) denote the standard deviation of $P_{x,y,z}$, where typical
detector counts reach $5\times10^5$; Solid lines are numerical fits
\cite{21}. (g) Spin-resolved spectra obtained from the $y$ component
spin polarization data. The non-Lorentzian lineshape of the
$I_y^{\uparrow}$ and $I_y^{\downarrow}$ curves and their non-exact
merger at large $|k_{x}|$ is due to the time evolution of the
surface band dispersion, which is the dominant source of statistical
uncertainty. a.u., arbitrary units. (h) Fitted values of the spin
polarization vector \textbf{P} = ($S_x$,$S_y$,$S_z$) are
(sin(90$^{\circ}$)cos(-95$^{\circ}$),
sin(90$^{\circ}$)sin(-95$^{\circ}$), cos(90$^{\circ}$)) for
electrons with +$k_x$ and (sin(86$^{\circ}$)cos(85$^{\circ}$),
sin(86$^{\circ}$)sin(85$^{\circ}$), cos(86$^{\circ}$)) for electrons
with -$k_x$, which demonstrates the topological helicity of the
spin-Dirac cone. The angular uncertainties are of order
$\pm$10$^{\circ}$ and the magnitude uncertainty is of order
$\pm$0.15.

\vspace{1cm}

\textbf{Figure 2 : Tuning the bulk Fermi level through systematic
bulk charge compensation monitored through systematic transport and
ARPES measurements.} (a) Resistivity at T = 4 K measured for samples
of Bi$_2$Se$_3$ that are bulk electron doped due to varying
concentrations of Se vacancies \cite{16} ($x$) or bulk hole doped
through Ca/Bi substitution ($\delta$). These are compared to
analogous values for topological insulators Bi$_{0.9}$Sb$_{0.1}$
(black square, arrowed; \cite{13}) and
Bi$_{1.9933}$Sn$_{0.0067}$Te$_3$ (purple triangle, arrowed). The
stoichiometric Bi$_2$Se$_3$ (Bi$_{1.9975}$Ca$_{0.0025}$Se$_3$) is
found to be the most insulating of these topological insulators. In
Bi$_{2-\delta}$Ca$_{\delta}$Se$_3$, bulk resistivity in excess of 75
m$\Omega$cm is possible, which will be shown elsewhere. The bulk
insulating state in Bi$_{0.9}$Sb$_{0.1}$ \cite{13} is intrinsic and
not due to disorder which will also be shown elsewhere.
Bi$_{1.9975}$Ca$_{0.0025}$Se$_3$ is known to be most metallic-like
among the three classes studied so far. (b) Hall carrier density of
the same samples determined using Hall measurements. Symbols colored
red (blue) represent n- (p-) type behavior. (c) ARPES band
dispersion images of Bi$_{2-\delta}$Ca$_{\delta}$Se$_3$ (111)
through collected within 20 minutes after cleavage for $\delta$ = 0,
(d) $\delta$ = 0.0025 and (e) $\delta$ = 0.01. Panels (f) through
(h) show the corresponding momentum distributions curves. Red lines
are guides to the eye. Panels (i) through (k) show the schematic
downward evolution of $E_F$ with increasing Ca content. The occupied
bulk conduction band (BCB) and bulk valence band (BVB) states are
shaded dark, and the occupied surface states (SS) are colored red.
(l) Typical ARPES band dispersion image of panels (c) through (e)
taken around 18 hours after cleavage and (m) its corresponding
momentum distribution curves. (n) Schematic of the surface band
bending process that is responsible for the observed downward shift
in energies over time. Vac., vacuum.

\vspace{1cm}

\textbf{Figure 3: Tuning the density of helical Dirac electrons to
the spin-degenerate Kramers point and topological transport regime.}
(a) A high resolution ARPES mapping of the surface Fermi surface
(FS) near $\bar{\Gamma}$ of Bi$_{2-\delta}$Ca$_{\delta}$Se$_3$
(111). The diffuse intensity within the ring originates from the
bulk-surface resonance state \cite{15}. (b) The FS after 0.1
Langmuir (L) of NO$_2$ is dosed, showing that the resonance state is
removed. (c) The FS after a 2 L dosage, which achieves the Dirac
charge neutrality point. (d) High resolution ARPES surface band
dispersions through after an NO$_2$ dosage of 0 L, (e) 0.01 L, (f)
0.1 L, (g) 0.5 L, (h) 1 L and (i) 2 L. The arrows denote the spin
polarization of the bands. We note that due to an increasing level
of surface disorder with NO$_2$ adsorption, the measured spectra
become progressively more diffuse and the total photoemission
intensity from the buried Bi$_{2-\delta}$Ca$_{\delta}$Se$_3$ surface
is gradually reduced.

\vspace{1cm}

\textbf{Figure 4 : Topological order of the nodal helical Dirac
ground state at 300 K.} (a) Typical ARPES intensity map of the
Bi$_2$(Se/Te)$_3$ class collected at $E_F$ spanning several
Brillouin zones. (b) Energy distribution curves of the valence bands
of Bi$_{2-\delta}$Ca$_{\delta}$Se$_3$ taken at T = 10 K, 200 K and
300 K. The peaks around -4 eV and -7.5 eV come from NO$_2$
adsorption (SI). The intensity of these NO$_2$ core level peaks do
not change over this temperature range, indicating no measurable
NO$_2$ desorption during the heating process. (c) ARPES intensity
map of the surface state band dispersion of
Bi$_{2-\delta}$Ca$_{\delta}$Se$_3$ (111) after a 2 L dosage of
NO$_2$ is applied at T = 10 K, which is taken at 300 K and (d) 10 K.
(e) Schematic of the surface and bulk electronic structure of
Bi$_{2-\delta}$Ca$_{\delta}$Se$_3$.NO$_2$, tuned to the topological
transport regime. (f) Angle-integrated intensity near (red) shows a
linear trend. The inset shows the expected density of states (DOS)
of a helical Dirac cone, which is 1/2 that of a graphene Dirac cone
due to its single spin degeneracy. (g) Schematic of the chiral Dirac
fermion ground state of graphene, which exhibits spin-degenerate
Dirac cones that intersect away from the Kramers' points. (h)
Schematic of the helical Dirac fermion ground state of
Bi$_{2-\delta}$Ca$_{\delta}$Se$_3$.NO$_2$, which exhibits a
spin-polarized Dirac cone that intersects at a Kramers' point and
guarantees a $\nu_0$ = 1 topological quantum number for the ground
state.

\newpage

\begin{figure}
\includegraphics[scale=0.6,clip=true, viewport=-0.0in 0in 11.0in 5.5in]{SpinBi2Te3v11ARR_CM}
\caption{\label{fig:Fig1} Hsieh, Xia, Qian $et$ $al.$, NATURE 460,
1101 (2009).}
\end{figure}

\newpage

\begin{figure}
\includegraphics[scale=0.65,clip=true, viewport=0.1in 0in 11.3in 6.0in]{bulk_doping9ARR_CM}
\caption{\label{fig:Fig2} Hsieh, Xia, Qian $et$ $al.$, NATURE 460,
1101 (2009).}
\end{figure}

\newpage

\begin{figure}
\includegraphics[scale=0.65,clip=true, viewport=-0.0in 0in 6.8in 6.8in]{BiSeNO2v3ARR_CM}
\caption{\label{fig:Fig3} Hsieh, Xia, Qian $et$ $al.$, NATURE 460,
1101 (2009).}
\end{figure}

\newpage

\begin{figure}
\includegraphics[scale=0.55,clip=true, viewport=-0.0in 0in 11.2in 6.0in]{summary2ARR_CM}
\caption{\label{fig:Fig4} Hsieh, Xia, Qian $et$ $al.$, NATURE 460,
1101 (2009).}
\end{figure}

\end{document}